\documentclass[a4paper]{EslabStyle}
\usepackage{epsfig}

\begin{document}

\title{Evolved Super Star Clusters in M82}

\author{R.  de Grijs\inst{1} \and R.\,W.  O'Connell\inst{1} \and J.\,S. 
Gallagher, {\sc iii}\inst{2}} \institute{Astronomy Department,
University of Virginia, P.O.  Box 3818, Charlottesville, VA 22903, USA
\and Astronomy Department, University of Wisconsin, 475 North Charter
Street, Madison, WI 53706.  USA}

\maketitle 

\begin{abstract}
We present high-resolution optical and near-infrared {\sl HST}
observations of two adjacent regions in the fossil starburst region in
M82, B1 and B2.  \\
Age estimates date the cluster population in the fossil starburst
between $\sim 2 \times 10^8$ and $\sim 10 \times 10^9$ years, assuming
solar metallicity.  The star cluster population in B2 is more heavily
affected by internal extinction than that in B1.  Although our cluster
size estimates indicate that they are gravitationally bound, and not
unlike Galactic globular clusters, their luminosity distribution is
significantly fainter than both super star cluster luminosity functions
(LFs) and the Galactic globular cluster LF.  This makes them unlikely
proto-globular clusters, since their luminosities will fade further
before they are of similar age as the Galactic globular cluster
population. \\
If the compact H$\alpha$-bright sources in M82 B are Type II supernova
remnants (SNRs), they set an upper limit to the end of the starburst in
region ``B2,'' about 500 pc from the galaxy's core, of $\sim 50$ Myr. 
Region ``B1,'' about 1000 pc from the core, lacks good SNR candidates
and is evidently somewhat older.  This suggests star formation in the
galaxy has propagated inward toward the present-day intense starburst
core. 
\end{abstract}

\section{Introduction}
  
\begin{figure*}
\centerline{\epsfig{file=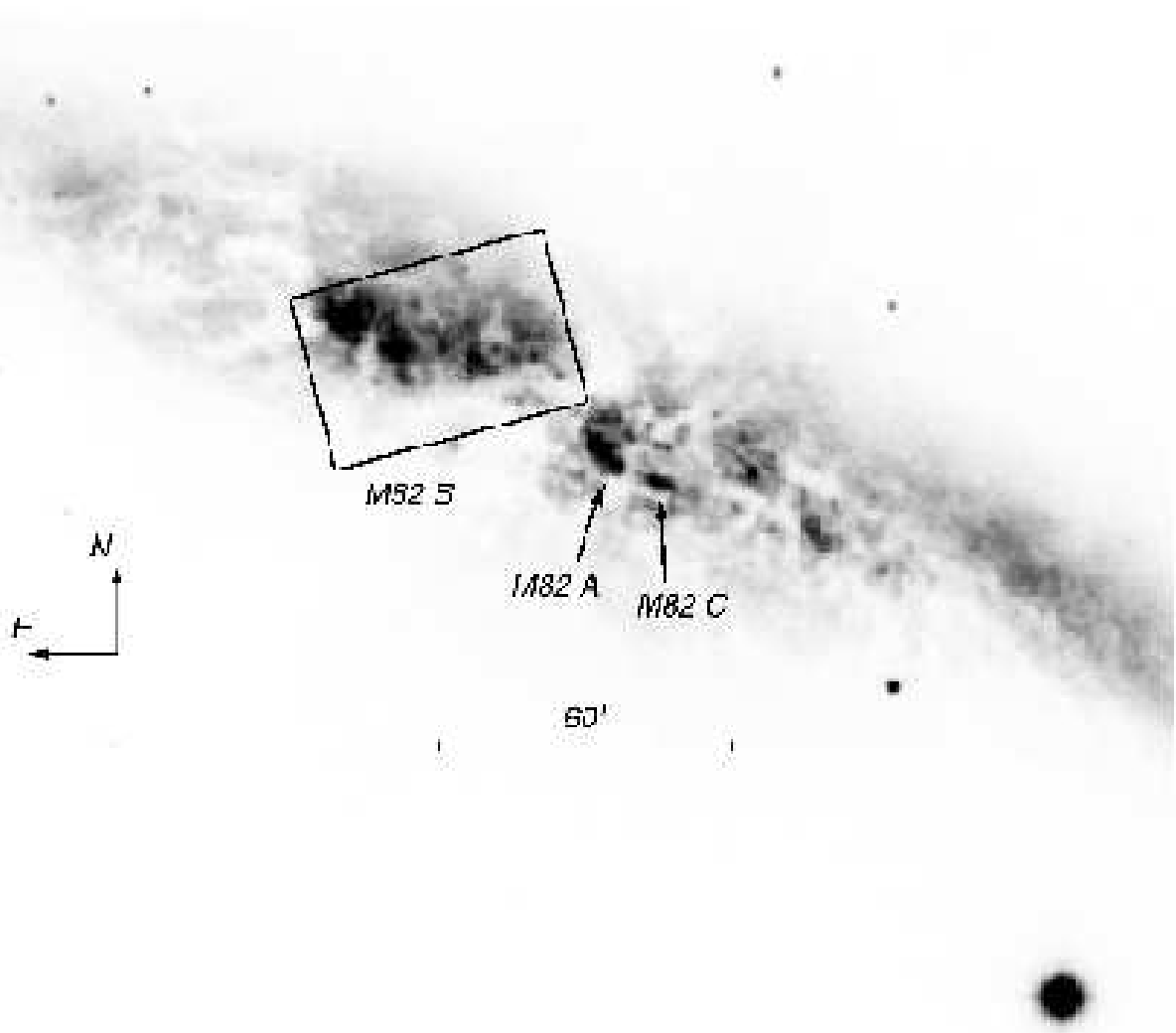}}
\caption{\label{m82abc.fig} A Palomar 5-m plate of M82 taken by Sandage
in the {\it B} band (20 minutes, seeing $\le 1''$, cf.  O'Connell et al. 
1995), identifying the regions we discuss.  The image is oriented north
up.  60$''$ corresponds to 1050 pc.  The regions of active star
formation, M82 A and C, are indicated.  The box labeled ``M82 B''
indicates the field of our H$\alpha$ observations; region B1 corresponds
to the eastern section of the box, east of B2.}
\end{figure*}

Observations at all wavelengths from radio to X-rays are consistent with
a scenario that tidal interactions between M82 and another member galaxy
of the M81 group have channeled large amounts of gas into the central
regions of M82 during the last several 100 Myr (e.g., Telesco 1988,
Rieke et al.  1993).  This has induced a starburst which has continued
for up to about 50 Myr.  All of the bright radio and infrared sources
associated with the active starburst are confined to the galaxy's
center, lying within a radius of $\sim 500$ pc, and corresponding
spatially with bright optical structures, labeled M82 A, C, and E in
O'Connell \& Mangano (1978, OM78), see Fig. \ref{m82abc.fig}. 

However, evidence exists that this is not the only major starburst
episode to have occurred in M82.  A region at about 1 kpc NE from the
galactic center, M82 B (cf.  OM78), has exactly the properties one would
predict for a fossil starburst with a similar amplitude to the active
burst.  Marcum \& O'Connell (1996) find a sharp main-sequence cut-off of
the composite stellar system in M82 B, corresponding to an age of $\sim$
100--200 Myr and an average extinction of $A_V \sim 0.6$ mag.  Region A,
on the other hand, is only consistent with a very young population
($\sim 5$ Myr) and is more heavily affected by internal extinction ($A_V
\sim 2.2$ mag).  By extrapolating region B's surface brightness ($\mu_V
\sim 16.5$ mag arcsec$^{-2}$, after correction for foreground
extinction; OM78) back to an age of 10 Myr we estimate that its surface
brightness was $\sim 2$ magnitudes brighter (cf.  the Bruzual \& Charlot
[1996, BC96] stellar population models), similar to that presently
observed in the active starburst. 

{\sl Hubble Space Telescope (HST)} imaging of the bright central regions
of M82 (A and C) resolved these into a swarm of young star cluster
candidates, with a FWHM of $\sim 3.5$ pc ($0.''2$) and mean $L_V \sim 4
\times 10^6 L_\odot$ (O'Connell et al.  1994, 1995), brighter than any
globular cluster in the Local Group. 

This is the nearest rich system of such objects; such ``super star
clusters'' (SSCs) have also been discovered with {\sl HST} in other
interacting and amorphous systems, and in dwarf and starburst galaxies
(e.g., Holtzman et al.  1992, Whitmore et al.  1993, O'Connell et al. 
1994, Conti et al.  1996, Ho 1997, Carlson et al.  1998, Watson et al. 
1998, among others).  Their diameters, luminosities, and -- in several
cases -- masses are consistent with these being young {\it globular}
clusters formed as a result of recent gas flows (e.g., van den Bergh
1995, Meurer 1995, Ho \& Filippenko 1996).  It is possible that most of
the star formation in starbursts takes place in the form of such
concentrated clusters.  Our observations of M82 do not reveal similar
cluster formation outside the active and the fossil starburst regions. 

Under the assumption that region B is indeed a fossil starburst site, it
is expected that it originally contained a complement of luminous
clusters similar to that now observed in region A.  The combination of
observations of both the active and the fossil starburst sites in M82
therefore provides a unique physical environment for the study of the
stellar and dynamical evolution of these star cluster systems. 

\section{{\sl HST} observations of the M82 central region}

The fossil starburst region, M82 B, was observed on September 15, 1997,
with both {\sl WFPC2} and {\sl NICMOS} on board the {\sl HST}.  We
imaged two adjacent $\sim35''$ square fields (Planetary Camera [PC]
field of view, $0.''0455$ pix$^{-1}$) in the M82 B region in the F439W,
F555W and F814W passbands, with total integration times of 4400s, 2500s
and 2200s, respectively, for region ``B1'', and 4100s, 3100s and 2200s,
respectively, for region ``B2''.  These observations were obtained using
four exposures per filter, covering a large range in integration times
to facilitate the removal of cosmic ray events.  The F439W, F555W and
F814W filters have roughly similar characteristics to the
Johnson-Cousins broad-band {\it B, V} and {\it I} filters, respectively. 

In the near-infrared (NIR) we chose to use {\sl NICMOS} Camera-2
($0.''075$ pix$^{-1}$), which provided the best compromise of resolution
and field of view.  We acquired 4 partially spatially overlapping
exposures in both the F110W and F160W filters (approximately similar to
the Bessell {\it J} and {\it H} filters, respectively) in a tiled
pattern; the integrations, of 512s each, were taken in MULTIACCUM mode
to preserve dynamic range and to correct for cosmic rays. 

\section{Selection procedure}

\subsection{Selection of real sources}

Unfortunately, the separation of real sources from artifacts in M82 B is
problematic, due to significant small-scale variations in the amplitude
of the background emission, which are largely caused by the highly
variable extinction.  For this reason, we cannot use standard unsharp
masking techniques to remove this background, since this produces
significant residual emission along the dust features. 

An initial visual examination of the multi-passband observations
revealed a multitude of faint point sources that become increasingly
obvious with increasing wavelength.  To include in our source selection
a maximum number of real and a minimum amount of spurious sources (due
to, e.g., dust features, weak cosmic rays, or poisson noise in regions
of high surface brightness or at the CCD edges), we decided to cross
correlate source lists obtained in individual passbands.  We performed
extensive checks to find the best selection criteria and thus to
minimize the effects introduced by artifacts on the one hand and the
exclusion of either very red or very blue sources on the other. 

We chose our detection thresholds such that the number of candidate
sources selected from the images in all passbands were comparable, of
order 4000.  Then, we cross correlated the source lists obtained in the
individual passbands.  Finally, we determined which combination of
passbands resulted in the optimal matching of sources detected in both
the blue and the NIR passbands.  This source selection procedure led us
to conclude that a final source list obtained from the cross correlation
of the candidate sources detected in the F555W and F814W filters would
contain the most representative fraction of the M82 B-region source
population. 

However, an initial visual examination of the cross-correlated {\it V}
and {\it I}-band sources showed that the automated detection routine had
returned many artifacts that were clearly associated with dust features. 
It also showed that the quality of the resulting sources for photometric
follow-up was highly variable.  Therefore, we decided to check the
reality of all cross-correlated sources by examining them visually, in
both passbands, at the same time classifying them in terms of contrast,
sharpness, and the presence of nearby neighbors.  Moreover, we added
(and verified) the $\sim$ 30\% of the sources in each field that were
missed by the automated detection routine, but were clearly real
sources, and were recognized as such by the visual examination of both
the {\it V} and the {\it I}-band images. 

The final source lists thus obtained contain 737 and 642 verified
sources in M82 B1 and B2, respectively. 

One should realize, however, that our {\it I}-band-to-NIR selection bias
will lead to rather conservative age estimates due to the older mean
ages of the sources emitting in these wavelength regimes: we visually
examined the distribution of our verified source population superposed
on the {\it B}-band images, and concluded that, by following the source
selection procedure outlined above, we missed less than $\sim 20$ of the
bluest (real) sources in each field due to this selection effect.  The
less complex galactic background at shorter (i.e., {\it V} band)
wavelengths helped to filter out artifacts of the data, the majority of
the resolved {\it stellar} population that becomes increasingly dominant
at longer wavelengths (see de Grijs et al.  2000b), and features due to
the background emission.  An important concern for the subsequent
analysis of our source sample is that we are only sampling the outer and
possibly slightly older surface of the M82 B region, while active star
formation, obscured by the ubiquitous dust component, may still be
ongoing inside this region. 

\subsection{Control fields}

We estimated the completeness of our object list by randomly and
uniformly adding 500 synthetic point sources of input magnitudes between
20.0 and 25.0 mag to the observed images in both the F555W and the F814W
filters.  The PSFs of the synthetic point sources were obtained from
observational PSFs, and scaled to the desired magnitudes.  The effects
of crowding in our simulated star fields are small: only $\sim$ 1--2\%
of the simulated objects were not retrieved due to crowding or overlap
of adjacent sources. 

For the uniformly distributed simulated sources, we established the 50\%
completeness limits at F555W as $> 23.1$ and 23.3 mag (for B1 and B2,
respectively) and at F814W as $> 23.0$ mag. 

In addition to these synthetic star fields, we used our $\omega$ Cen
{\sl HST} observations in the F555W passband, obtained as part of
program GO-6053, to verify the reduction procedures and the accuracy of
our photometry. 

\subsection{Separating star clusters from stars}

We based the distinction between stars and more extended sources in the
M82 B fields on the statistical differences between the size
characteristics of the populations of real sources in these fields and
those of the stars in the $\omega$ Cen control field.  To do so, we added
a scaled version of the $\omega$ Cen field to the M82 B fields, such that
the output magnitudes of the majority of the $\omega$ Cen stars were in
the same magnitude range as those of the M82 B sources, in the latter
reference frame. 

\begin{figure}[!h]
\centerline{\epsfig{file=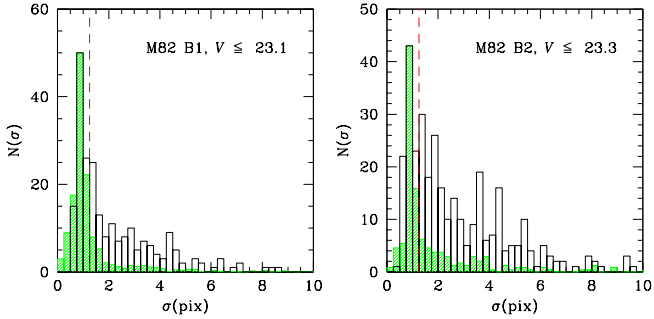,width=22pc}}
\vspace{-4cm}
\caption{\label{selection.fig}Distribution of $\sigma_{\rm Gaussian}$
for B1 and B2 (open histograms).  All sources down to and including the
50\% completeness limits have been included.  The shaded histograms are
the corresponding size distributions for the $\omega$ Cen field,
synthetically added to the M82 B frames and scaled to the peak of the
M82 B histograms.  We have indicated the size limit adopted to
distinguish between stars and star clusters by the dashed line at
$\sigma_{\rm Gaussian} = 1.25$ pixels.}
\end{figure}

We determined characteristic sizes of both our M82 B verified sources
and the $\omega$ Cen stars, synthetically added to the M82 frames, using
a Gaussian fitting routine.  Although the true luminosity profiles of
the star clusters in M82 B may differ from Gaussians (de Grijs et al. 
1999, 2000b), this method allows us to distinguish between compact and
extended sources.  A comparison between the distribution of
characteristic stellar sizes from the globular cluster field with those
of the verified sources in M82 B revealed that the population of
extended sources in M82 B is well-represented by sources with
$\sigma_{\rm Gaussian} \ge 1.25$ pixels, see Fig.  \ref{selection.fig}. 
Therefore, in the following we will consider those sources with
$\sigma_{\rm Gaussian} \ge 1.25$ and {\it V}-band magnitude brighter
than the 50\% completeness limit to be part of our verified cluster
samples.  They contain 128 and 218 cluster candidates in B1 and B2,
respectively. 

\section{Cluster colors, ages and extinction}
\label{populations.sect}

\begin{figure}
\centerline{\epsfig{file=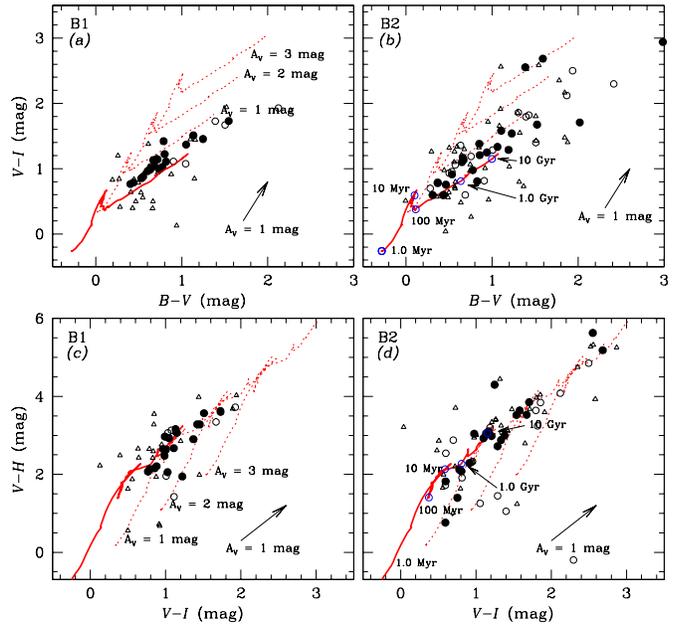,width=22pc}}
\caption{\label{CCs.fig}Optical and NIR color-color diagrams for the
cluster candidates in M82 B1 and B2, divided in magnitude bins (roughly
corresponding to quality bins); solid dots: $V \le 21.0$; open circles:
$21.0 < V \le 22.0$; open triangles: $22.0 < V \le$ 50\% completeness
limit.  The reddening vectors are shown as arrows; the effects of
reddening on unreddened evolutionary tracks of an instantaneous burst
stellar population (BC96, full drawn lines) are indicated by the dotted
tracks.  We have only included those sources that were not significantly
affected by nearby neighbors or highly variable backgrounds.}
\end{figure}

A close examination of the color-color diagrams of the cluster
populations in either field (Fig.  \ref{CCs.fig}) reveals that M82 B2 is
more heavily affected by internal extinction than region B1 (assuming
similar, Galactic foreground extinction), as is evidenced by the larger
spread of the data points along the direction of the reddening vectors.
Most of the redder cluster colors observed, in particular in region B2,
can be achieved by up to 2(--3) mag of visual extinction (assuming a
similar extinction law as in the Galaxy and a foreground screen
extinction geometry).

Employing the BC96 initial burst models (solid lines in Fig. 
\ref{CCs.fig}) yields ages between $\sim 2 \times 10^8$ and $\sim 10
\times 10^9$ yr for region B1, and a slightly smaller lower age limit
for B2, assuming solar metallicity.  Near-solar metallicity should be a
reasonable match to the young objects in M82 (e.g., Gallagher \& Smith
1999); Fritze-v.  Alvensleben \& Gerhard (1994) also show that chemical
evolution models indicate that young clusters should have metallicities
$Z \sim 0.3 - 1.0 Z_\odot$.  Other than the slightly bluer envelope in
M82 B2 with respect to B1, indicating more recent star formation in B2,
the representative age range and upper age limits inferred from the
panels in Fig.  \ref{CCs.fig} appears to be similar for either region. 
The upper age limit of $\sim 10$ billion years may be indicative of the
onset of the gravitational interaction with M81 (or another galaxy in
the same group) that produced the star clusters presently seen in the
post-starburst region M82 B. 

\section{Comparison with Galactic globular clusters}

\subsection{Sizes}

We need a good estimate of the star clusters' linear sizes to determine
if they are as compact as (Galactic old) globular clusters (e.g.,
Whitmore \& Schweizer 1995, among others), or more diffuse, like open
clusters or stellar (OB) associations (e.g., van den Bergh 1995, but see
Meurer 1995), which may be less tightly bound and may therefore dissolve
over the clusters' lifetimes (e.g., van den Bergh 1995, Miller et al. 
1997).  Any stellar associations in our sample that are gravitationally
unbound, and with net expansion velocities of order 1 km s$^{-1}$
(Blaauw 1964) will -- over the allowed lifetimes of $\sim 0.2$ to $\sim
10$ billion years -- have expanded to radii of $\sim 200$ pc to $\sim 10$
kpc.  On the other hand, sources with effective radii of order 10 pc and
lifetimes of a few hundred Myr are many crossing times (of typically a
few Myr) old and therefore likely gravitationally bound globular
clusters (e.g., Schweizer et al.  1996). 

The majority of the extended sources in the M82 B regions are
characterized by effective radii of $2.34 \le R_{\rm eff} \le 8.4$ pc
(based on Gaussian fits of the stellar radial luminosity behavior, but
see Holtzman et al.  1996 and de Grijs et al.  1999), with a wing
extending to larger sizes.  Due to our size selection threshold at
$\sigma_{\rm Gaussian} = 1.25$ pixels, we are likely only probing the
more extended star clusters, since we cannot distinguish star clusters
with smaller effective radii from stars.  Therefore, these effective
radii are consistent with size estimates obtained from {\sl HST} imaging
for the majority of young globular clusters and cluster systems in other
galaxies (e.g., O'Connell et al.  1994, Barth et al.  1995, Holtzman et
al.  1996, Schweizer et al.  1996, Miller et al.  1997, Whitmore et al. 
1999) as well as in the Galaxy.  The Galactic globular cluster
population is characterized by a median effective radius of $\sim 3$ pc
(van den Bergh et al.  1991, Djorgovski 1993, van den Bergh 1995), and a
total range from 0.7 to 20 pc. 

Thus, it appears that the population of star cluster candidates in the
M82 B regions resembles the Galactic globular cluster population at a
younger age.  We point out that in order to become old globular
clusters, the sizes of these objects should not change significantly
over their lifetimes.  Additional support for an insignificant evolution
of effective radii over the first $\sim 1$ Gyr is provided by our
observations of the star cluster candidates in the actively star forming
region, A, in the center of M82 (O'Connell et al.  1995).  The recently
formed young cluster candidates in M82 A are found to have typical
effective radii of $\sim 3.5$ pc, which makes them closely resemble the
slightly evolved populations in M82 B1 and B2. 

In de Grijs et al.  (2000b) we show that similar conclusions are reached
if we take, e.g., modified Hubble laws (among others) as representative
star cluster profiles. 

\subsection{The cluster luminosity function}

\begin{figure}
\centerline{\epsfig{file=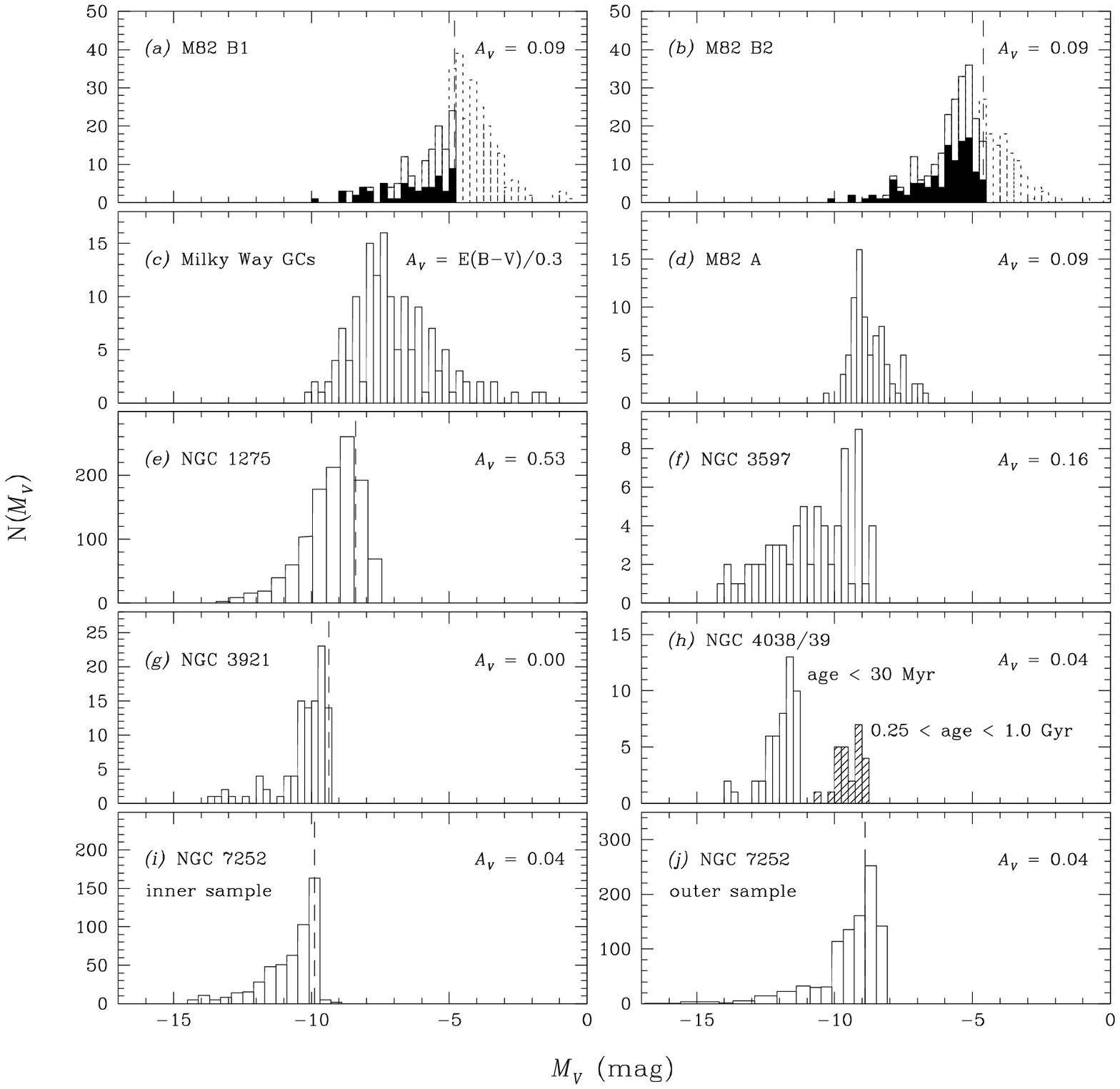,width=22pc}}
\caption{\label{absv.fig} Cluster luminosity functions of the M82 B
regions, compared to the Galactic globular cluster population, and
recently published {\it young} star cluster populations in other nearby
galaxies, based on {\sl HST} observations.  The CLFs have been corrected
for Galactic foreground extinction (Burstein \& Heiles 1984); extinction
corrections towards Milky Way globular clusters were adopted from Harris
(1996).  Where available we have indicated the completeness limits by
the dashed lines.  The absolute magnitude determinations are based on
``true'' distance estimates where possible; in all other cases distance
moduli were calculated for $H_0 = 50$ km s$^{-1}$ Mpc$^{-1}$.  {\it (a)}
and {\it (b)} M82 B1 and B2 (this paper; only sources brighter than the
completeness limits and with $\sigma_{\rm Gaussian} \ge 1.25$ pixels are
included, the shaded CLFs represent the subsamples with $\sigma_{\rm
Gaussian} \ge 3$ pixels); {\it (c)} Milky Way globular clusters (Harris
1996); {\it (d)} M82 A (O'Connell et al.  1995); {\it (e)} NGC 1275
(Carlson et al.  1998); {\it (f)} NGC 3597 (Holtzman et al.  1996); {\it
(g)} NGC 3921 (Schweizer et al.  1996); {\it (h)} NGC 4038/39 (Whitmore
et al.  1999) -- open CLF: young clusters ($< 30$ Myr); shaded CLF:
intermediate-age clusters ($0.25 - 1.0$ Gyr); {\it (i)} and {\it (j)}
NGC 7252 (Whitmore et al.  1993; Miller et al.  1997), inner ($r < 6''$)
and outer ($r > 6''$) sample, respectively.}
\end{figure}

Fig.  \ref{absv.fig} shows the absolute {\it V}-band cluster luminosity
functions (CLFs) for B1 and B2; for comparison, we have also included
the CLFs of the globular cluster system in the Galaxy and recently
published ({\sl HST}-based) CLFs for nearby galaxies that contain a
statistically significant number of young (super) star clusters (where
applicable, we used $H_0 = 50$ km s$^{-1}$ Mpc$^{-1}$ to obtain absolute
magnitudes), including M82 A (O'Connell et al.  1995).  Where available,
we have indicated the completeness limits by dashed lines. 

From this figure, it is immediately clear that the star clusters in M82
B are significantly fainter than the SSCs recently detected by {\sl HST}
in other galaxies.  They are also significantly fainter than the
majority of the Galactic globular cluster population, which must have
faded by several magnitudes compared to the young SSCs, due to stellar
evolution (cf.  BC96).  Considering the younger ages of the M82 B star
clusters with respect to the Galactic globular cluster population, it is
unlikely that the M82 B CLF will resemble the Galactic globular cluster
LF after evolving to similar ages: stellar and star cluster evolution
indicates an additional fading with time, the amount of which depends on
the current ages of the M82 star clusters (cf.  BC96). 

\section{Supernova Remnants}

To supplement our ongoing broad-band imaging program on the SSCs in M82
B, we extracted from the {\sl HST} archive H$\alpha$+continuum
observations of the central regions of M82, taken through the F656N
narrow-band filter (March 16, 1997, program GO-6826; and September 12,
1995, GO-5957).  Since we lacked a suitable narrow-band continuum image
near the bandpass of the F656N filter, we created a pseudo-continuum
image from our co-registered (and essentially emission-line-free) {\it
V} and {\it I}-band {\sl WFPC2} images, by linearly interpolating the
continuum fluxes to the mean wavelength of the F656N filter.  We
subtracted the pseudo-continuum image thus constructed from the
H$\alpha$+continuum image to obtain an image containing pure line
emission (see de Grijs et al.  2000a for a detailed technical
description). 

Close examination of the continuum subtracted H$\alpha$ image of M82 B
shows that region B1 has few compact H$\alpha$ sources.  However, B2,
located closer to the active starburst core, is brighter in H$\alpha$,
due to both compact sources and diffuse emission, although still at a
surface brightness $\sim 20 \times$ lower than in the active starburst. 
Most of the identified H$\alpha$ sources in B2 have H$\alpha$
brightnesses significantly above the norm for the exterior regions of
the galaxy, especially considering the excess extinction in B2 (see
Sect.  \ref{populations.sect}).  The brightest H$\alpha$ sources could
not be cross correlated with any of the verified continuum sources
described in the previous sections, however.  The brightest object in B2
resolves to a ``string of pearls'' of discrete sources and is
conspicuously located adjacent to the strong central dust lane which
separates region B from the starburst core. 

One expects to find two types of compact H$\alpha$ sources in a galaxy
like M82.  H{\sc ii} regions will exist around young star clusters with
ages $\le 10$ Myr by virtue of the presence of ionizing O- or early
B-type stars.  However, there should always be a significant continuum
source associated with such very young H{\sc ii} regions, even if there
is considerable extinction in the vicinity.  Type II supernovae (SNe)
can also produce compact H$\alpha$ remnants.  These will often be
associated with their parent star clusters, but in many cases there may
be no well-defined compact continuum source.  Since Type II SNe can
occur up to 50 Myr after a star formation event, the associated cluster
may have faded or dynamically expanded enough to be inconspicuous
against the bright background of the galaxy.  Alternatively, the parent
star of the SNR could have been ejected from the cluster or could have
formed initially in the lower density field.  In fact, earlier studies
of resolved starbursts suggest that 80-90\% of the bright stellar
population resides in a diffuse component outside of compact clusters
(e.g., Meurer et al.  1995, O'Connell et al.  1995). 

The identified M82 B H$\alpha$ sources fall into two wide luminosity
ranges: those with $L ({\rm H}\alpha) < 9 \times 10^{35}\, {\rm erg}\,  
{\rm s}^{-1}$ and those with $L({\rm H}\alpha) > 14 \times 10^{35}\,  
{\rm erg}\, {\rm s}^{-1}$.  Neither group has the properties expected   
for normal H{\sc ii} regions in the disk of a late-type galaxy.  Recent
samples of normal H{\sc ii} regions in spiral and irregular galaxies
have been compiled by, among others, Kennicutt, Edgar \& Hodge (1989), 
Bresolin \& Kennicutt (1997), and Youngblood \& Hunter (1999).  The
brightest H$\alpha$ source in M82 B has $L(H\alpha) < 10^{37}\, {\rm
erg}\, {\rm s}^{-1}$.  Although normal galaxies contain many H{\sc ii}
regions with luminosities in this range, they invariably also have much
brighter sources, with luminosities up to 10$^{38-39} {\rm erg}\, {\rm
s}^{-1}$.  The largest M82 B source is smaller than 20 pc in diameter,  
whereas typical diameters for normal disk H{\sc ii} regions are 30--100
pc.

Instead, we believe that the 10 sources in the more luminous group are
good candidates for SNRs.  Six of these have only faint counterparts in
the continuum passbands.  Some of the sources show evidence of limb
brightening, as might be expected for older SNRs.  Further
(circumstantial) evidence for their SNR nature is discussed in de Grijs
et al.  (2000a). 

\section{Propagating star formation?}

The presence of SNRs in the post-starburst region of M82 can help to set
limits on its star formation history.  The last SNe in a quenched
starburst region would occur at a time comparable to the longest
lifetime of an SN progenitor after the end of the starburst activity. 
Following Iben \& Laughlin (1989) and Hansen \& Kawaler (1994), the time
$t$ spent between the zero-age main sequence and planetary nebula phase
by an $8 {\rm M}_\odot$ progenitor star, which is generally adopted as a
lower limit for Type II SNe (e.g., Kennicutt 1984), corresponds to $t
\sim 35-55$ Myr.  Type Ia SNe, which involve lower mass stars in binary
systems, can occur much later, but one expects these to be more
uniformly distributed over the galaxy's surface, not concentrated near
regions of recent star formation.  The radiative lifetimes of the shell
SNRs in their later phases are short in this context.  They are limited
by the expansion velocities of their shells. 

Therefore, if our candidates are indeed SNRs they suggest an upper limit
to the end of the starburst event in region B2 of $\sim 50$ Myr.  The
absence of SNR candidates in B1 indicates it is older.  The ages derived
from our analysis of the color-color diagrams of the M82 B1 and B2
clusters (Sect.  \ref{populations.sect}, de Grijs et al.  2000b), and
those of spectral synthesis dating of region B1 ($\sim$ 100--200 Myr,
Marcum \& O'Connell 1996) are consistent with this.  There is then the
following progression of ages with distance from the center of the
current starburst: B1 ($\ge$ 100 Myr, $r \sim 1000$ pc), B2 ($\le$ 50
Myr, $r \sim 500$ pc), and the present core (active for $\le 20$ Myr, $r
< 250$ pc), and this suggests that intense star-forming activity in M82
has propagated inward toward the present starburst core during the past
100--200 Myr. 

\begin{acknowledgements}

This work is based on observations with the NASA/ESA {\sl Hubble Space
Telescope}, obtained at the Space Telescope Science Institute, which is
operated by the Association of Universities for Research in Astronomy
(AURA), Inc., under NASA contract NAS 5-26555.  We acknowledge funding
from NASA grants NAG 5-3428 and NAG 5-6403. 

\end{acknowledgements}

\end{document}